\documentclass[%
 reprint,
superscriptaddress,
 amsmath,amssymb,
 aps,
]{revtex4-2}

\usepackage{graphicx}
\usepackage{dcolumn}
\usepackage{bm}

\begin{document}

\preprint{APS/123-QED}

\title{Synthesizing $2h/e^2$ resistance plateau at the first Landau level confined in a quantum point contact}

\author{M.~H.~Fauzi}
\email{moha042@brin.go.id}
\affiliation{Research Center for Quantum Physics, National Research and Innovation Agency, South Tangerang 15314, Indonesia and Research Collaboration Center for Quantum Technology $2.0$, Bandung 40132, Indonesia}

\author{K.~Nakagawara}
\affiliation{Department of Physics, Tohoku University, Sendai 980-8578, Japan}

\author{K.~Hashimoto}
\affiliation{Department of Physics, Tohoku University, Sendai 980-8578, Japan}
\affiliation{Center for Science and Innovation in Spintronics, Tohoku University, Sendai 980-8577, Japan}

\author{N. Shibata}
\affiliation{Department of Physics, Tohoku University, Sendai 980-8578, Japan}

\author{Y. Hirayama}
\email{yoshiro.hirayama.d6@tohoku.ac.jp}
\affiliation{Center for Science and Innovation in Spintronics, Tohoku University, Sendai 980-8577, Japan}
\affiliation{Takasaki Advanced Radiation Research Institute, QST, 1233 Watanuki-machi, Takasaki, 370-1292 Gunma, Japan}

\date{\today}

\begin{abstract}
A comprehensive understanding of quantum Hall edge transmission, especially a hole-conjugate of a Laughlin state such as a $2/3$ state, is critical for advancing fundamental quantum Hall physics and enhancing the design of quantum Hall edge interferometry. In this study, we report a robust intermediate $2h/e^2$ resistance quantization in a quantum point contact (QPC) when the bulk is set at the fractional filling $2/3$ quantum Hall state. Our results suggest the occurrence of two equilibration processes. First, the co-propagating $1/3$ edges moving along a soft QPC arm confining potential fully equilibrate and act as a single $2/3$ edge mode. Second, the $2/3$ edge mode is further equilibrated with an integer $1$ edge mode formed in the QPC. The complete mixing between them results in a diagonal resistance value quantized at $2h/e^2$. Similar processes occur for a bulk filling $5/3$, leading to an intermediate $(2/3)h/e^2$ resistance quantization.

\end{abstract}

\maketitle


\section{Introduction}

\begin{figure}[t]
\begin{center}    
\centering
\includegraphics[width=\linewidth]{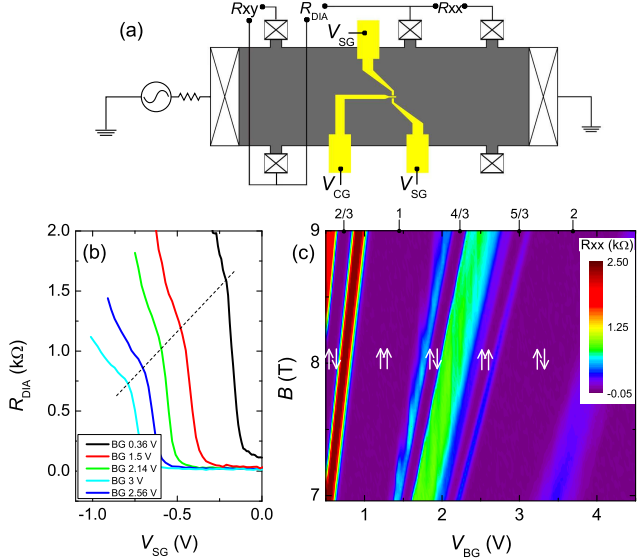}
\end{center}
\caption{(a) schematic of device and measurement setup. A pair of split gates defines a quantum point contact (QPC). An additional center gate in between the split gates is used to control the density inside the QPC. The length of the split gates is $500$ nm with the gap between them being $500$ nm. The width of the center gate is $300$ nm. (b) Zero-field diagonal resistance as a function of $V_{\mathrm{SG}}$ measured at three different $V_{\mathrm{BG}}$ values. The dashed line corresponds to the threshold voltage to deplete electron density underneath the split gates. The split gates are equally biased throughout the measurements. (c) 2D map of bulk transport $R_{\mathrm{xx}}$ measured as a function of $V_{\mathrm{BG}}$ and magnetic field $B$. Several notable filling factors and the corresponding spin-states are indicated.}
\label{Fig01} 
\end{figure}

When a two-dimensional electron gas (2DEG) is placed in a strong magnetic field at low temperature, its Hall resistance can be quantized into $(1/\nu)h/e^2$ (or $\nu e^2/h$ in conductance). The Fractional Quantum Hall Effects (FQHEs) occur when the filling factor $\nu$ takes a fractional number. The FQHEs exhibit several remarkable properties, such as fractional charge excitations \cite{Laughlin1983}, fractional or anyonic statistics \cite{Wilczek1982}, and chiral Luttinger edge modes \cite{Wen1992}. The nature of Luttinger liquid is manifested in the power-law behavior of the tunneling conductance, which has been experimentally verified \cite{Chang1996, Grayson1998}. While the original Laughlin states, $1/(2m + 1)$ with $m$ being an integer, carry a single chiral edge mode as described by Wen \cite{WenPRL1990, WenPRB1991}, other fractional states derived from the Laughlin states, such as in the Haldane-Halperin hierarchy \cite{Haldane1983, Halperin1984}, can be reconstructed and may carry multiple propagating edge modes \cite{MacDonald1990}. A prominent example is the $\nu = 2/3$ fractional state, the hole-conjugate of Laughlin $\nu = 1/3$ state, where several edge models have been contested to date \cite{MacDonald1990, Kane1994, Meir1994, Nosiglia2018}.

The reconstructed edge states for the fractional $2/3$ depend on the profile of the edge confining potential. For a sharp confining potential, such as the edge of a Hall bar, MacDonald has proposed that the $2/3$ edge is reconstructed into two distinct counter-propagating states: a downstream integer $1$ edge state and an upstream fractional $-1/3$ edge state \cite{MacDonald1990}. An ideal B\"uttiker two-terminal conductance for the MacDonald edge picture should result in $(4/3)e^2/h$ quantized conductance \cite{Buttiker1988}; however, $(2/3)e^2/h$ is usually observed. A theory proposed by Kane-Fisher-Polchinski (KFP) suggests that the two terminal conductance quantization can yield $(2/3)e^2/h$ due to the formation of a single downstream $2/3$ charge mode and an upstream neutral mode in the presence of disorders and interactions between the edges \cite{Kane1994}. This edge model has been verified through a recent experiment by Cohen \textit{et al.} \cite{Cohen2019}.  Conversely, for a soft confining potential, Meir has proposed that two co-propagating downstream fractional $1/3$ edge channels are more favorable to form \cite{Meir1994}. Indeed, in a QPC with a soft confining potential, an intermediate quantized conductance $(1/3)e^2/h$ is observed \cite{Bhattacharyya2019, Nakamura2022}, consistent with the Meir edge picture.


Recently, Fu \textit{et al} \cite{Fu2019} reported a $G_{\mathrm{QPC}} = (3/2)e^2/h$ plateau when the bulk filling is set to $\nu_{\mathrm{bulk}} = 5/3$ in an ultra-high mobility constriction and Hayafuchi \textit{et al} \cite{Hayafuchi2022} reported a similar plateau in a conventional QPC with a center gate. Yan \textit{et al} \cite{Yan2022} conducted an extended experiment over a wider bulk filling factor and suggested an important role of an elevated density in the constricted region. Removing the outer integer $1$ edge channel from $G_{\mathrm{QPC}} = (3/2)e^2/h$ and $\nu_{\mathrm{bulk}} = 5/3$ should result in $G_{\mathrm{QPC}} = (1/2)e^2/h$ and $\nu_{\mathrm{bulk}} = 2/3$. Motivated by this point, we perform similar measurements for our QPC devices with the center gate at the $\nu_{\mathrm{bulk}} = 2/3$ and $\nu_{\mathrm{bulk}} = 5/3$.

Recently, Nakamura et al. \cite{Nakamura2022} reported an interesting observation of an intermediate half-conductance plateau $G_{\mathrm{QPC}} = (1/2)e^2/h$ at a finite dc bias in their QPC at the $\nu_{\mathrm{bulk}} = 2/3$ by utilizing a specially designed GaAs/AlGaAs heterostructures. They focused on the sharpness of the confining potential profile. With their specially designed structure, they are able to induce a sharp QPC confining potential where the MacDonald edge picture is favorable to form. They explain the intermediate half-conductance plateau as being due to the full reflection of the inner counter-propagating $-1/3$ edge mode. Our conventional QPC has a soft confining potential where the Meir edge picture is more favorable. Under the aforementioned conditions, we observe a clear intermediate $G_{\mathrm{QPC}} = (1/2)e^2/h$ plateau without the dc bias.

\section{Results and Discussion}

Our measurement is carried out in an 18-nm-wide GaAs/AlGaAs quantum well. The 2DEG is located $185$ nm from the surface. The wafer is processed into a $100$ $\mu$m Hall bar. The 2DEG electron density at low temperatures is controlled by a back gate $V_{\mathrm{BG}}$. The device and measurement setup are schematically displayed in Fig. \ref{Fig01}(a). A source-drain AC current of $1$ nA is fed into the device and all the resistive tensor components ($R_{\mathrm{xx}}$, $R_{\mathrm{xy}}$, and $R_{\mathrm{DIA}}$) are measured by a standard lock-in amplifier technique at a frequency of $17.3$ Hz. The QPC is defined by applying a negative bias voltage to a pair of spit gates ($V_{\mathrm{SG}}$). Additional center metal gate ($V_{\mathrm{CG}}$) put in between the split gates is used to control the QPC electron density\cite{Hayafuchi2022}, as well as the QPC confinement potential\cite{Chou1993, Lee2006, Maeda2016}.

Fig. \ref{Fig01}(b) displays the basic operation of our QPC at zero magnetic fields under various 2DEG densities controlled by the back gate voltage $V_{\mathrm{BG}}$. The threshold voltage, at which the electron density underneath the split gates is completely depleted, scales linearly with $V_{\mathrm{BG}}$ (indicated by the dashed line in Fig. \ref{Fig01}(b)). This is consistent with that of a parallel plate capacitor model\cite{Davies1995}. By applying a more negative bias to $V_{\mathrm{SG}}$, we can observe a series of quantized conductance steps, similar to the previous report in Ref. \cite{Hayafuchi2022}.

Fig. \ref{Fig01}(c) displays a 2D map of bulk longitudinal resistance $R_{\mathrm{xx}}$ as a function of $V_{\mathrm{BG}}$ and magnetic field $B$. We can attribute the spin polarization for each quantum Hall state observed in Fig. \ref{Fig01}(c), in particular for the fractional $2/3$ and $5/3$ states. The spin for the fractional $2/3$ state, within the magnetic field window shown in Fig. \ref{Fig01}(c), is unpolarized as we do not observe a spin transition to spin-polarized. We expect the spin transition to occur at a higher magnetic field for a narrow quantum well owing to the large Coulomb interaction \cite{Fauzi2014}. The fractional $5/3$ state is a particle-hole conjugate of the $\nu = 1/3$ Laughlin state \cite{Laughlin1983}, and therefore is always spin-polarized.

\begin{figure}[t!]
\begin{center}    
\centering
\includegraphics[width=\linewidth]{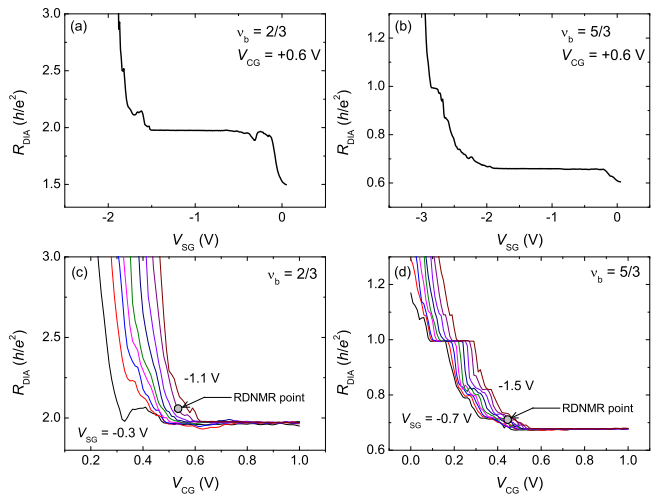}
\end{center}
\caption{(a) Diagonal resistance $R_{\mathrm{DIA}}$ measured as a function of $V_{\mathrm{SG}}$ with the bulk filling factor set to $\nu_{\mathrm{b}} = 2/3$. (b) $R_{\mathrm{DIA}}$ measured as a function of $V_{\mathrm{SG}}$ with the bulk filling factor set to $\nu_{\mathrm{b}} = 5/3$. An extended resistance plateau is visible in both cases. (c)-(d) Similar to the case in panel (a)-(b), but the $V_{\mathrm{CG}}$ is swept instead at several $V_{\mathrm{SG}}$ values at an interval of $0.1$ V. All the data are measured at $100$ mK and $7$ T.}
\label{RDIA} 
\end{figure}

Fig. \ref{RDIA}(a) displays the transport through the QPC with the bulk fractional filling set to $\nu_{\mathrm{b}} = 2/3$. The center gate is biased with a positive voltage $V_{\mathrm{CG}} = +0.6$ V, resulting in a density inside the QPC that is higher than the bulk. As the $V_{\mathrm{SG}}$ is swept toward a negative voltage and passes the threshold voltage, the diagonal resistance gets quantized to within $2\%$ of $R_{\mathrm{DIA}} = 2h/e^2$. The plateau is extended over a wide range of $V_{\mathrm{SG}}$ before the channel finally pinches off with no other intermediate plateaus observed in between. This exceptionally long extended plateau suggests the existence of a stable edge channel formed inside the QPC. The plateau persists over a wide gate voltage and temperature range to approximately $300$ mK \cite{section2}. No intermediate $3h/e^2$ resistance plateau is observed, as in the usual case, presumably a non-linear confining potential created by the center gate destabilizes the formation of $\nu = 1/3$ state in our QPC \cite{Ito2021}. No finite dc bias is required to stabilize the $2h/e^2$ plateau, as in Ref. \cite{Nakamura2022}, suggesting that the inter-edge back-scattering in the QPC is highly suppressed.

We then examine the influence of applying positive center gate bias $V_{\mathrm{CG}}$ on transport properties. Applying positive $V_{\mathrm{CG}}$ increases the electron density in the QPC. $V_{\mathrm{CG}}$ is swept at several fix $V_{\mathrm{SG}}$ bias voltages from $-0.3$ down to $-1.1$ V with an interval of $0.1$ V as displayed in Fig. \ref{RDIA}(c). To deplete the electron underneath the split gate, a threshold voltage of $-0.21$ V is required (see the black curve for $V_{\mathrm{BG}} = 0.36$ V in Fig. \ref{Fig01}(b)) for the QPC to always be defined. A pronounce deep in the $R_{\mathrm{DIA}}$ at $V_{\mathrm{SG}} = -0.3$ V before reaching the plateau, shown in Fig. \ref{RDIA}(c), is indicative of interaction between an edge channel running outside the QPC and a classical skipping orbit formed inside the QPC \cite{Haug1989a, Haug_1993}. The dip disappears when a more negative bias voltage is applied to $V_{\mathrm{SG}}$ because the effective channel length increases. All the curves are merged toward $2h/e^2$ with the increase in $V_{\mathrm{CG}}$. This observation complements the result shown in Fig. \ref{RDIA}(a), but more importantly, it gives us a solid idea that the density inside the QPC is higher when the $2h/e^2$ quantization occurs.

\begin{figure}[t]
\begin{center}    
\centering
\includegraphics[width=\linewidth]{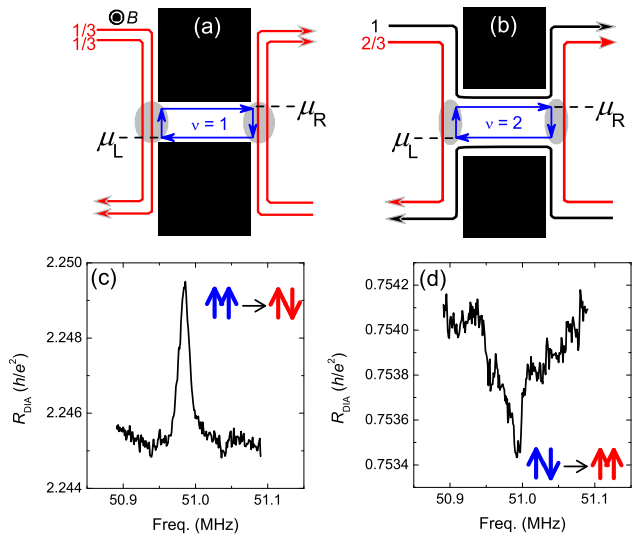}
\end{center}
\caption{Schematic of edge structure when the inner edge $2/3$ mixes with an integer (a) $1$ or (b) $2$ in the QPC, leading to the resistance quantization at $2h/e^2$ and $(2/3)h/e^2$, respectively. The grey circle region indicate the equilibration. $\mu_{\mathrm{L}}$ and $\mu_{\mathrm{R}}$ denotes the chemical potential at which the edges are fully equilibrated. (c)-(d) RDNMR spectra recorded in the QPC at the bulk filling $2/3$ and $5/3$ with increasing rf frequency. The inset shows the spin-flip scattering inducing nuclear spin polarization. The RDNMR spectra are carried out near the $2h/e^2$ and $(2/3)h/e^2$, respectively.}
\label{Edgemodel} 
\end{figure}

We perform similar measurements for a bulk filling factor $5/3$ as displayed in Fig. \ref{RDIA}(b) at a magnetic field of $7$ T. The filling factor $5/3$ is supposed to have an edge structure similar to that of the filling factor $2/3$, but with an additional outer integer $1$ edge mode. Certainly, we observe an intermediate long plateau at $(2/3)h/e^2$ followed by a short intermediate $h/e^2$ plateau before the channel pinches off, in line with our previous study using a wider GaAs quantum well and shorter QPC geometry size \cite{Hayafuchi2022} and in Ref. \cite{Fu2019, Yan2022}. Sweeping $V_{\mathrm{CG}}$ at a fix $V_{\mathrm{SG}}$ as shown in Fig. \ref{RDIA}(d) yields quantitatively a similar result where all the curves are merged to the $(2/3)h/e^2$ plateau but with the $h/e^2$ plateau being more well-developed. Interestingly, we also observe a plateau-like structure at around $(0.8)h/e^2$ and $(0.7)h/e^2$ between $(2/3)h/e^2$ and $h/e^2$ plateau which are absent in Fig. \ref{RDIA}(b).

Now, let us discuss the model to explain $2h/e^2$ and $(2/3)h/e^2$ resistance quantization. We consider the co-propagating $1/3$ fractional edges running along the soft edge potential equilibrating with the integer $1$ edge at the entrance and exit of the point contact, as indicated by the grey circle in Fig. \ref{Edgemodel} (a). The fully equilibrated point, where the chemical potential becomes the same, can be modeled as an Ohmic contact \cite{Yan2022} and we label them as $\mu_{\mathrm{L}}$ and $\mu_{\mathrm{R}}$. Considering the equilibration length for $\nu = 2/3$ at $6.5$ T is about $8$ $\mu$m \cite{Lin2019} and the QPC arm is much longer than $8$ $\mu$m, it is reasonable to assume that the co-propagating $1/3$ edge modes are in the incoherent regime (fully equilibrated) \cite{Christian2020} before entering the QPC and mixing with the integer $1$ edge mode. We can presume the co-propagating $1/3$ edge mode as a single $2/3$ edge mode when it mixes with the integer $1$ edge mode. Considering these two equilibration processes, we can express the formula similar to that derived by Yan \textit{et al.} in Ref. \cite{Yan2022}
\begin{equation}
    R_{\rm{DIA}} = \frac{h}{e^2}/\left(\frac{\nu_{\rm{b}} - i}{2 - (\nu_{\rm{b}} - i)} + i\right)
\end{equation}
Plugging $\nu_{\rm{b}} = 2/3$ and $i = 0$ yield $2h/e^2$ resistance quantization, in agreement with the observed value. We can get $(2/3)h/e^2$ value as well for the bulk filling factor $5/3$ with $i = 1$ since we have an additional outer integer $1$ edge mode as indicated in Fig. \ref{Edgemodel}(b). It is important to note that the actual path where the inter-mode edge mixing happens may not follow a straight line, as drawn in the schematic. However, the actual path may follow a more complex route owing to the geometry of the center gate.


We can infer the edge spin polarization by spin-flip scattering-induced dynamic nuclear polarization (DNP) and resistively-detected nuclear magnetic resonance (RDNMR) \cite{Machida2002}. DNP requires a momentum-conserving spin flip-flop process between two edges with different spin polarities. In other words, no DNP occurs and hence no RDNMR signal if the edges have similar spin polarities. Furthermore, the polarity of nuclear spin polarization is determined by the direction of the spin-flip process \cite{Fauzi2017}.

Since the filling factor $2/3$ at a magnetic field of $7$ T is a spin-unpolarized state as noted in Fig. \ref{Fig01}(c) and the filling factor $1$ is a spin-polarized state, we expect the spin-flip process to occur in the QPC with the direction indicated in the inset of Fig. \ref{Edgemodel}(c). The flip should polarize the nuclei in the direction parallel to the external magnetic field. RDNMR measurements are carried out near the plateau at a point shown in Fig. \ref{RDIA}(c)-(d). We sweep the rf frequency around $^{75}$As nuclei as displayed in Fig. \ref{Edgemodel}(c) and observe a peak in the diagonal resistance at a frequency of $50.986$ MHz. On the other hand, when we tune the bulk to the filling factor $5/3$ and the QPC to the filling factor less than $2$, we expect a reversal in the RDNMR signal. This is because the filling factor $2$ is a spin-unpolarized state while the filling factor $5/3$ is a spin-polarized state. Indeed, we observe a reversal in the RDNMR spectrum as displayed in Fig. \ref{Edgemodel}(d) with a resonance frequency occurring at $50.993$ MHz. Although the fractional $2/3$ and the inner fractional $5/3$ are supposed to have the same edge mode, however, our RDNMR measurements have revealed that they have different spin structures. 




In summary, we have synthesized $2h/e^2$ and $(2/3)h/e^2$ resistance quantization in the QPC within $2\%$ deviation. We have offered a model to explain the observed quantization based on edges mixing between the co-propagating $1/3$ edges and integer $1$ edge formed inside the QPC. We have also confirmed the spin polarization of these edges through current-induced DNP and RDNMR measurements, which is consistent with the proposed model to explain the $2h/e^2$ and $(2/3)h/e^2$ quantization. Understanding the interactions of edges in a QPC is crucial for improving the design of quantum Hall interferometry and quantum Hall-based quantum circuitry.

\acknowledgements
We thank K. Muraki and NTT Basic Research Laboratory for supplying a high-quality GaAs wafer used in this experiment. M. H. F acknowledges Graduate School in Spintronics, Tohoku University for financial support at Tohoku University where the experiments were conducted. This work was supported in part by KAKENHI Grant Number JP18H01811 (Y. H), JP20H05660 (K. H, Y. H), JP22H01847  (K. H, Y. H, N. S), and JP19K03708 (N. S).

\section*{Author Contributions}
M. H. F and Y. H. conceived the experiments. M. H. F fabricated devices and conducted measurements with the help of K. N. and K.H. N.S. supported the theoretical parts.  M. H. F. wrote the manuscript along with the feedback from all co-authors. The project was supervised by Y. H.

\section*{Competing Interests} The authors have no competing financial interests to
declare.

%

\end{document}



\title{Supplementary for synthesizing $2h/e^2$ resistance plateau at the first Landau level confined in a quantum point contact}







\maketitle


\begin{figure}[t]
\begin{center}    
\centering
\includegraphics[width=0.95\linewidth]{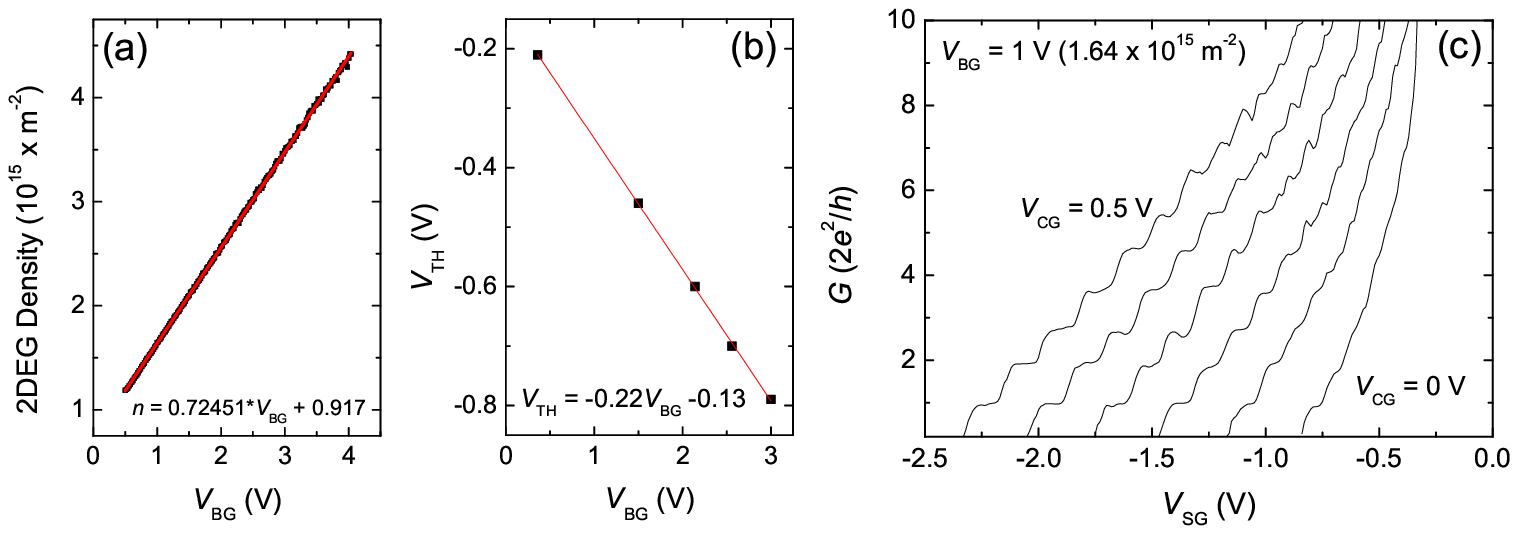}
\end{center}
\caption{(a) 2DEG density as a function of $V_{\mathrm{BG}}$. The 2DEG density is field-induced by applying $V_{\mathrm{BG}}$. The red line is a linear fit to the data from which the relation is obtained. (b) The threshold voltage ($V_{\mathrm{TH}}$) required to deplete the electron density underneath the split gates as a function of $V_{\mathrm{BG}}$. (c) Quantized conductance at zero magnetic fields taken at several $V_{\mathrm{CG}}$ bias voltages from $0$ to $0.5$ V with an interval of $0.1$ V. The split gates are biased equally.}
\label{FigS1} 
\end{figure}

\begin{figure*}[t]
\begin{center}    
\centering
\includegraphics[width=\linewidth]{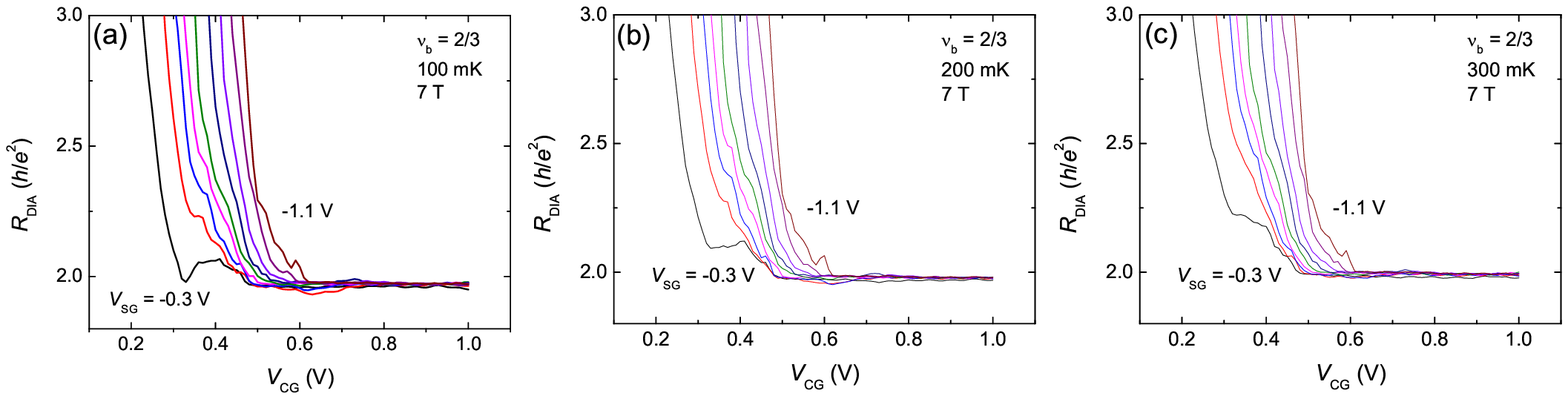}
\end{center}
\caption{Diagonal resistance $R_{\mathrm{DIA}}$ vs $V_{\mathrm{CG}}$ with the bulk filling factor set to $\nu_{\mathrm{b}} = 2/3$ taken at a temperature of (a) $100$ mK, (b) $200$ mK, and (c) $300$ mK.}
\label{FigS2} 
\end{figure*}

\begin{figure*}[t]
\begin{center}    
\centering
\includegraphics[width=\linewidth]{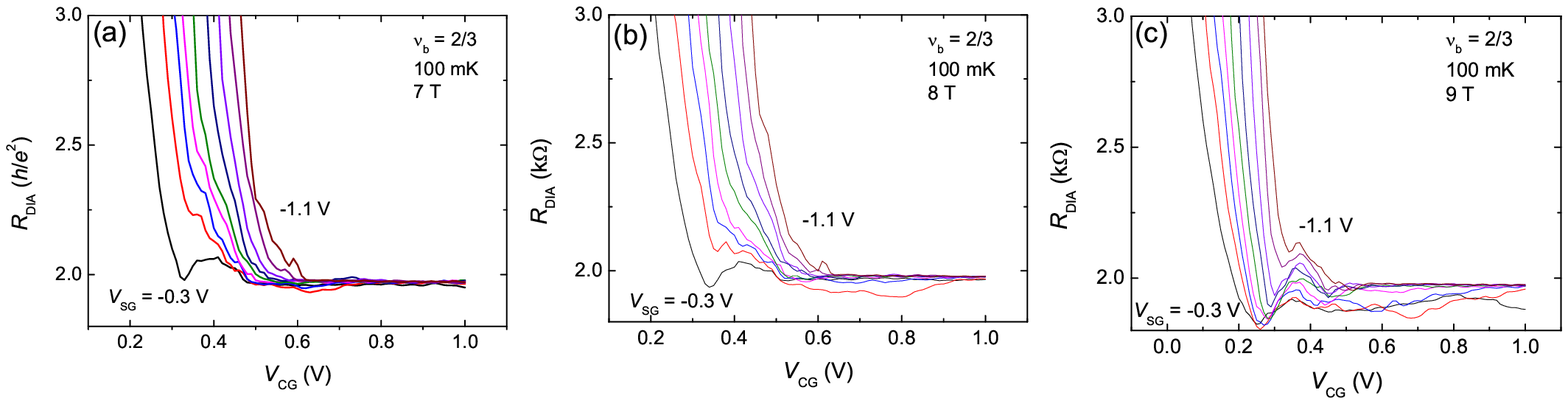}
\end{center}
\caption{Diagonal resistance $R_{\mathrm{DIA}}$ vs $V_{\mathrm{CG}}$ with the bulk filling factor set to $\nu_{\mathrm{b}} = 2/3$ taken at a magnetic field of (a) $7$ T, (b) $8$ T, and (c) $9$ T.}
\label{FigS3} 
\end{figure*}

\begin{figure}[t]
\begin{center}    
\centering
\includegraphics[width=0.4\linewidth]{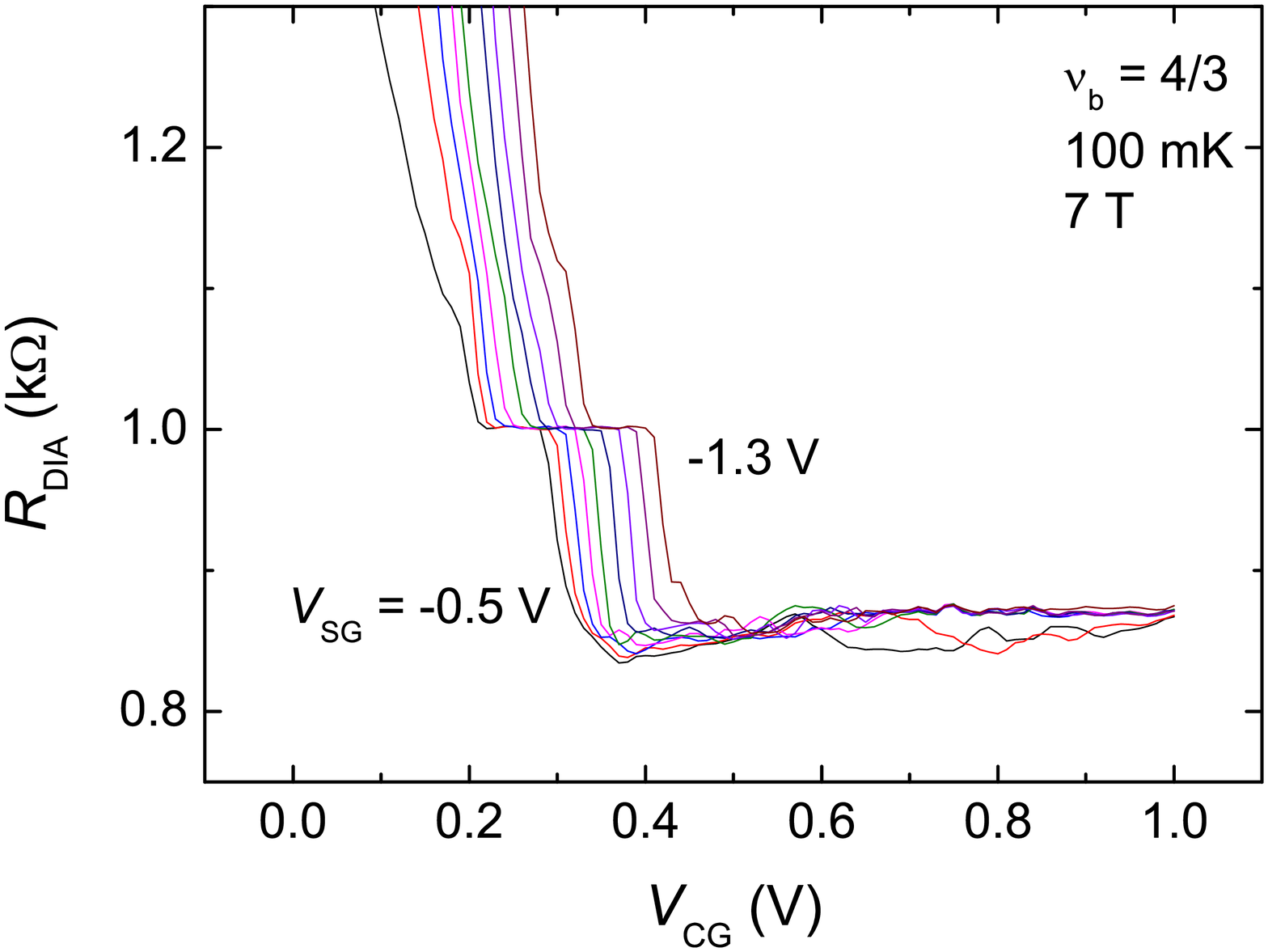}
\end{center}
\caption{Diagonal resistance $R_{\mathrm{DIA}}$ vs $V_{\mathrm{CG}}$ with the bulk filling factor set to $\nu_{\mathrm{b}} = 4/3$ taken at $7$ T and $100$ mK. $V_{\mathrm{SG}}$ is varied from $-0.5$ to $-1.3$ V at an interval of $0.1$ V.}
\label{FigS4} 
\end{figure}

\section{Device and Basic transport}

The Ohmic contact pads of our device are made of Ni/AuGe/Ni alloys, which were rapidly annealed at $390^0$ C for 1 minute in a hydrogen gas flow chamber. Our device has four metal gates in total, each with a specific function. The gate voltage $V_{BG}$ is used to control the density of the two-dimensional electron gas (2DEG) as shown in Fig. S1(a). The density of the 2DEG is found to scale linearly with $V_{BG}$, as expected. We can easily control the bulk filling factor by using $V_{BG}$ and magnetic field $B$. A pair of split gates ($V_{SG}$) and a center gate ($V_{CG}$) are used to control the transport properties of the quantum point contact (QPC). The threshold voltage required to deplete the electron density underneath the split metal gates under several back gate voltages is displayed in Fig. S\ref{FigS1}(b); it exhibits a linear dependence.

Fig. S\ref{FigS1}(c) shows the quantized conductance profile of our QPC as $V_{SG}$ is swept. We compare the profile at five $V_{SG}$ bias voltages. An increase in the center gate bias voltage results in more quantized steps in the conductance profile due to an increase in the subband spacing and enhanced screening \cite{Lee2006, Maeda2016}.


\section{Temperature dependence}

The $2h/e^2$ resistance quantization persists at least up to $300$ mK as displayed in Fig. S\ref{FigS2}. We do not observe a significant change in the resistance profile except the dip in the resistance at $V_{\mathrm{SG}} = -0.3$ V, before reaching the plateau, is reduced at higher temperatures. We expect the integer $1$ edge mode to be stably formed in the QPC even at elevated temperatures.

\section{Magnetic field dependence}

We compare the $2h/e^2$ resistance quantization at different magnetic fields as displayed in Fig. S\ref{FigS3}. The bulk filling factor is maintained at the filling $2/3$ by adjusting the electron density using $V_{\mathrm{BG}}$. The same $V_{\mathrm{SG}}$ bias voltage range is used for all cases in Fig. S\ref{FigS3}; however, the bias is always applied above the threshold voltage to deplete the electron density under the split metal gates. Besides a more pronounced deep in the resistance at a higher magnetic field, a more negative $V_{\mathrm{SG}}$ bias is required to stabilize the $2h/e^2$ resistance quantization.

\section{Other fractional - $5/6$ state}

The fractional $4/3$ carries two downstream integer $1$ and fractional $1/3$ edge modes. We expect the inner $1/3$ edge mode to fully equilibrate with the integer $1$ edge mode formed inside the QPC, resulting in a quantized resistance of $(5/6)h/e^2$, which is supported by the results shown in Fig. S\ref{FigS4}, which shows that the resistance quantization is close to $(5/6)h/e^2$ with a $4\%$ deviation. However, the deviation from the exact quantization is greater than that observed when mixing the fractional $2/3$ and integer $1$ edge modes.

The deviation in quantization cannot be attributed to a series resistance from the bulk, as the bulk fractional $4/3$ state is well-developed and the longitudinal resistance $R_{xx}$ approaches zero, as seen in Fig. 1(c) of the main text. Bulk back-scattering is highly suppressed in our $100$ $\mu$m wide Hall bar, unlike a narrower $30$ $\mu$m Hall bar used in a previous study \cite{Hayafuchi2022}. A possible explanation for the deviation could be greater inter-edge back-scattering within the QPC, as the electron density difference required to establish the $(5/6)h/e^2$ plateau is greater than it is to establish the $2h/e^2$ and $(2/3)h/e^2$ plateaus.

%